\documentclass[12pt,letterpaper]{scrartcl}
\usepackage{setspace}
\setkomafont{disposition}{\normalfont\bfseries}
\usepackage{hyperref}
\hypersetup{colorlinks=true,urlcolor=black,citecolor=blue}
\usepackage{amsfonts,amsmath,amssymb}
\usepackage{txfonts}
\usepackage{graphicx}
\usepackage{bm}
\usepackage{epstopdf}
\usepackage{verbatim}
\usepackage{xcolor}
\usepackage[margin=1.0in]{geometry}
\usepackage{units}
\usepackage[T1]{fontenc}
\usepackage[latin9]{inputenc}
\usepackage{amsbsy}
\usepackage{esint}
\usepackage[super,comma,sort&compress]{natbib}
\usepackage{lipsum}
\usepackage[font={small,it},labelfont=bf]{caption}
\usepackage{enumitem}
\usepackage{wrapfig}
\usepackage{caption}
\usepackage{sidecap}
\sidecaptionvpos{figure}{c}
\newcommand{\Dp}{\Delta p}

\newcommand{\simc}{\!\sim\!}
\newcommand\bb[1]{\mbox{\boldmath{$#1$}}}

\renewcommand{\bf}{\bfseries}
\renewcommand{\rm}{\mathrm}

\makeatletter
\renewcommand{\fnum@figure}{Fig.~\thefigure}
\makeatother

\begin{document}

\begin{center}
{\bf\huge The Material Properties of \\Weakly Collisional, High-$\bb{\beta}$ Plasmas} \\
\vspace{2ex}
{\bf\large White Paper for Plasma 2020 Decadal Survey} \\
\vspace{4ex}
Matthew Kunz \\
Dept.~of Astrophysical Sciences, Princeton University and \\ Princeton Plasma Physics Laboratory; \url{mkunz@princeton.edu} \\
\vspace{1ex}
and \\
\vspace{1ex}
Jonathan Squire \\
Dept.~of Physics, University of Otago, New Zealand;  \url{jonathan.squire@otago.ac.nz} \\
\vspace{1.5em}
February 2019 \\
\vspace{1.5em}
Co-authors: \\
\vspace{1em}
\begin{tabular}{ll}
    Steven Balbus & Dept.~of Physics, University of Oxford \\
    \mbox{} & New College, Oxford \\
    Stuart Bale & Dept.~of Physics, University of California, Berkeley \\
    \mbox{} & Space Sciences Laboratory at the University of California, Berkeley \\
    Christopher Chen & School of Physics and Astronomy, Queen Mary University of London\\
    Eugene Churazov & Max Planck Institute for Astrophysics, Garching \\
    \mbox{} & Space Research Institute (IKI), Moscow \\
    Steven Cowley & Dept.~of Astrophysical Sciences, Princeton University \\
    \mbox{} & Princeton Plasma Physics Laboratory \\
    Cary Forest & Dept.~of Physics, University of Wisconsin--Madison \\
    Charles Gammie & Dept.~of Astronomy, University of Illinois at Urbana-Champaign \\
    \mbox{} & Dept.~of Physics, University of Illinois at Urbana-Champaign \\
    Eliot Quataert & Dept.~of Astronomy, University of California, Berkeley \\
    \mbox{} & Dept.~of Physics, University of California, Berkeley \\
    Christopher Reynolds & Institute of Astronomy, University of Cambridge \\
    \mbox{} &  Sidney Sussex College, Cambridge \\
    Alexander Schekochihin & Rudolf Peierls Centre for Theoretical Physics, University of Oxford \\
    \mbox{} & Merton College, Oxford \\
    Lorenzo Sironi & Dept.~of Astronomy, Columbia University \\
    Anatoly Spitkovsky & Dept.~of Astrophysical Sciences, Princeton University \\
    James Stone & Dept.~of Astrophysical Sciences, Princeton University \\
    \mbox{} & Dept.~of Applied and Computational Mathematics, Princeton University\\
    Irina Zhuravleva & Dept.~of Astronomy and Astrophysics, University of Chicago \\
    Ellen Zweibel & Dept.~of Astronomy, University of Wisconsin--Madison \\
    \mbox{} & Dept.~of Physics, University of Wisconsin--Madison
\end{tabular}

\vspace{2ex}
(All authors and co-authors are listed alphabetically.)

\end{center}

\newpage

\subsection*{Introduction}

This white paper concerns the physics of weakly collisional, high-$\beta$ plasmas -- plasmas in which the thermal pressure $P$ dominates over the magnetic pressure ($\beta\equiv 8\pi P/B^{2}\gg{1}$, where $B$ is the magnetic-field strength), and in which the inter-particle collision time is comparable to the characteristic timescales of bulk motions. This state of matter, although widespread in the Universe, remains poorly understood: we lack a predictive theory for how it responds to perturbations, how it transports momentum and energy, and how it generates and amplifies magnetic fields. Such topics are foundational to the scientific study of plasmas, and are of intrinsic interest to those who regard plasma physics as a fundamental physics discipline. But these topics are also of extrinsic interest: addressing them directly informs upon our understanding of a wide variety of space and astrophysical systems, including accretion flows around supermassive black holes, the intracluster medium (ICM) between galaxies in clusters, and regions of the near-Earth solar wind.

An intriguing aspect of this state of matter concerns the relevance of the magnetic field, which -- although energetically subdominant -- influences the transport properties of the plasma by imparting directionality and new degrees of freedom to the system,
thereby influencing the large-scale dynamics. This causes fundamental differences between the dynamics of such plasmas and those of neutral gases, even in the limit of negligibly weak magnetic fields ($\beta\gg{1}$). Here we outline major scientific gaps in our understanding and the progress that is being made towards addressing them. After a brief description of some theoretical difficulties involved, we outline key application areas that are ripe for enhancing cross-disciplinary research opportunities between plasma physics and astrophysics. Following on this, we discuss recent progress in the field on a number of different fronts, providing guidance on how research priorities can be addressed over the next decade.

\begin{SCfigure}[][t]
  \captionsetup{format=plain}
    \includegraphics[width=0.5\textwidth]{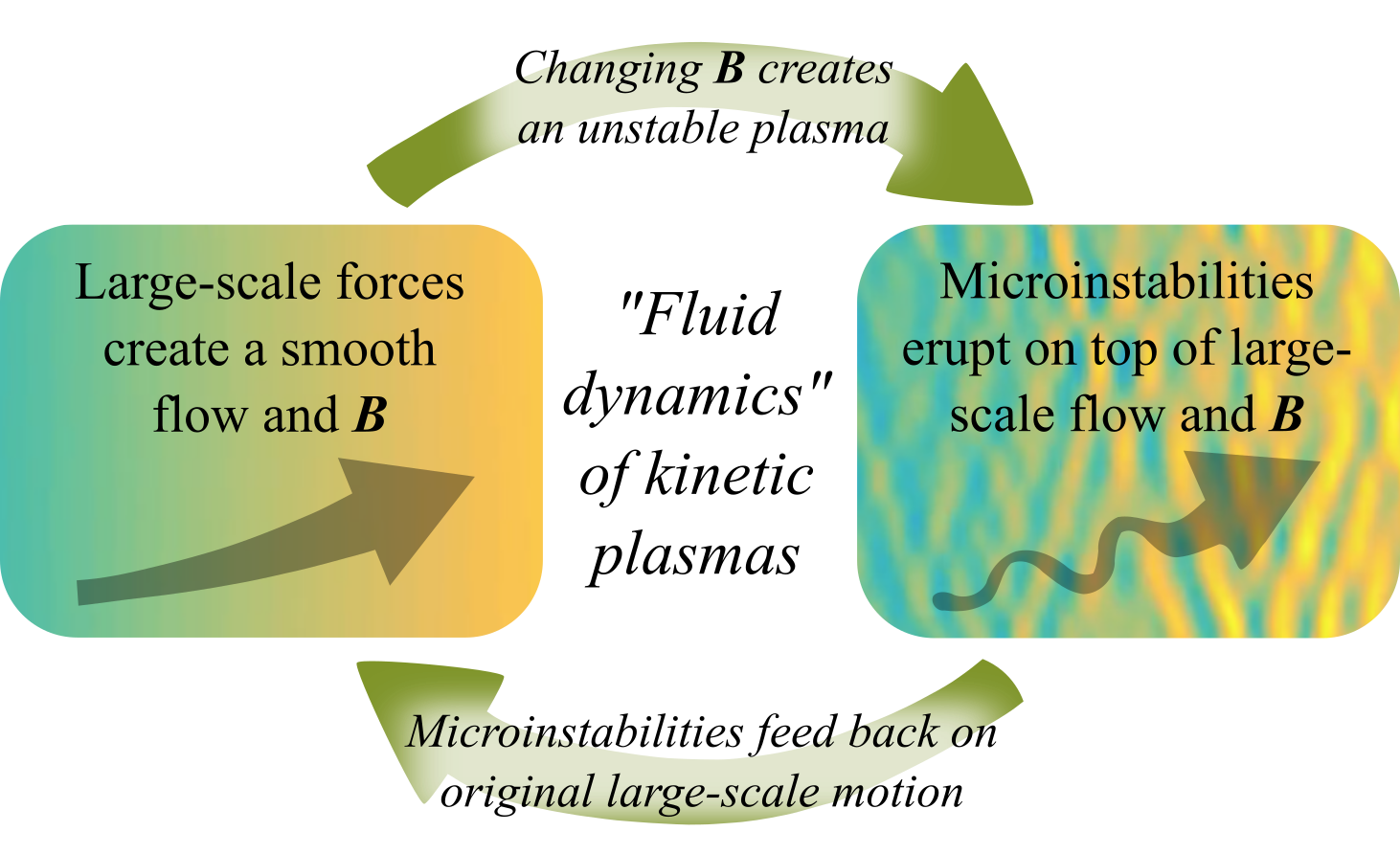}
    \caption{A diagram summarizing the difficulties involved in formulating a theory for the large-scale (``fluid'') dynamics of weakly collisional, high-$\beta$ plasmas. 
    Bulk changes in the plasma parameters (e.g., in $\bm{B}$) cause the smooth field of a macro-scale motion (left) to spontaneously erupt into gyroscale kinetic instabilities (right). These instabilities then modify the material properties of the plasma, feeding back on the large-scale fields and modifying the motions that caused them in the first place.}
    \label{fig:micros}
\end{SCfigure}

\vspace{-2ex}
\paragraph*{Theory: pressure anisotropy and kinetic instabilities.} A unique feature of weakly collisional, magnetized plasmas when $\beta\gtrsim{1}$ is their propensity to become unstable to kinetic gyroscale instabilities (``microinstabilities''). This stems from the near conservation of single-particle adiabatic invariants, most notably the magnetic moment $\mu\propto v_{\perp}^{2}/B$, where 
$v_{\perp}$ is the particle's perpendicular velocity. 
An increase/decrease in $B$ causes a commensurate increase/decrease in thermal pressure, but only 
in the direction perpendicular to the magnetic field. This pulls the plasma out of local thermodynamic equilibrium.
Microinstabilities are excited if the magnitude of the {pressure anisotropy} $\Dp\equiv p_{\perp}-p_{\|}$, where $p_\perp$ ($p_{\|}$) is the perpendicular (parallel) thermal pressure, exceeds ${\sim}B^2/4\pi$. Large field-aligned heat fluxes driven by temperature gradients can also drive such instabilities. At high $\beta$, this can occur for very small (order ${\sim}\beta^{-1}$) fluctuations, implying that even modest-amplitude, large-scale motions can destabilize the plasma on microscopic scales.\cite{Schekochihin2005,Squire2017a} In many astrophysical plasmas, these scales (e.g., the ion gyroscale) are far smaller than the system size (see Table \ref{tab:systems}), 
implying that these instabilities grow and saturate effectively instantaneously compared to the timescales of  large-scale motions.
Thus, we have a fluid whose properties are determined by the complicated 
nonlinear saturation of kinetic instabilities, which can provide an effective collisionality, influence heat fluxes, and modify the particle distribution 
function in complicated ways. This interaction between fluid and kinetic processes is summarized in Fig.~\ref{fig:micros}.

While our understanding of this interaction has progressed significantly in recent years, this field of study remains in its infancy. And while  computational methods can be used to study some problems in isolation, they remain too computationally expensive for direct application to many astrophysical problems of interest. We lack a set of simple fluid-like equations that captures the effects described above, a shortcoming that is impeding progress on multiple scientific frontiers.

\subsection*{Application areas}

Here we outline key topics in astrophysics and space physics for which a better understanding of weakly collisional, high-$\beta$ plasmas is needed for theoretical progress. These topics offer many cross-disciplinary opportunities where basic plasma physics can have a transformative impact.

\vspace{-2ex}
\paragraph{Magnetogenesis and dynamo.} {The origin and evolution of cosmic magnetism remains one of the most important outstanding questions of cosmology and astrophysics\cite{Kulsrud2008}: How were the first fields generated? Can plasma processes account for the strength and structure of the observed fields today? Answering these questions requires understanding both \emph{magnetogenesis} -- how magnetic fields came about in the first place -- and \emph{dynamo}, how fields are amplified and sustained by plasma motions. Hot, weakly collisional plasmas are thought to be key sites of cosmic magnetogenesis, suggesting that the themes presented above -- the relevance of field direction and the impact of kinetic instabilities -- play key roles in the origin of dynamically important magnetic fields. However, most theoretical work has employed the magnetohydrodynamic (MHD) model\cite{Brandenburg2005,Schekochihin2004}, 
often further assuming that fields grow without back-reacting on the plasma motions (the ``{kinematic}'' approximation). Although  recent studies\cite{Rincon2016,St-Onge2018} have started to explore dynamos affected by kinetic physics, our understanding of weakly collisional dynamos remains severely limited, representing a major gap in basic plasma-physics theory -- one which simultaneously touches on topics in kinetic turbulence, collisionless reconnection, non-thermal particle acceleration and diffusion, anomalous viscosity and resistivity, etc. Similarly, while we know of plasma processes and instabilities that  create magnetic fields\cite{Huntington2015}, most saturate with weak, tangled fields, and it remains unclear whether such fields were the seeds of those we observe today. We note that ``cosmic magnetism'' is one of five key science drivers of the upcoming Square Kilometer Array mega-project.}

\vspace{-2ex}
\paragraph{The intracluster medium (ICM).}
{The ICM is a hot, diffuse, X-ray-emitting plasma that resides in the deep gravitational potential well of galaxy clusters.\cite{Peterson2006} With measured $B \sim 1~\mu{\rm G}$ and kpc-scale mean-free paths, the ICM provides a
classic example of a weakly collisional, high-$\beta$ plasma.\cite{Carilli2002,Schekochihin2005} 
A fundamental, unanswered question in cluster physics is the ``cooling-flow'' problem, which concerns how the cores of clusters are thermally regulated in the presence of otherwise fast cooling through X-ray emission.\cite{Fabian1994} Plasma physics likely plays a fundamental role in answering this question: 
thermal transport (e.g., conduction\cite{Narayan2001}, convection\cite{Balbus2000,Quataert2008}) and heating (e.g., from active galactic nuclei\cite{Yang2016}) are dependent upon how large-scale plasma properties are influenced by kinetic plasma instabilities. 
While the baryonic content of clusters is dominated by the ICM, ${\sim}1/3$ of clusters also exhibit diffuse non-thermal emission (``radio halos'') from relativistic electrons in ${\sim}\mu{\rm G}$ intracluster magnetic fields.\cite{Ferrari2008} The origin of radio halos is a puzzle, as the short radiative lifetime of the emitting electrons implies their continuous {\em in situ} production or (re-)acceleration (e.g., by turbulence).\cite{Brunetti2014} Evaluating which mechanism is dominant requires a theory for how effectively cosmic rays are confined by self-induced streaming instabilities\cite{Kulsrud1969} 
and other high-$\beta$ plasma processes.
}
\begin{table}
\centering
\begin{tabular}{ || c  | c c c c c c c  ||  }
 \hline
 & $n\,(\mathrm{cm}^{-3})$ & $T \,(\mathrm{keV})$ & $B \,(\mu\mathrm{G})$ & $\beta$ & $L \,(\mathrm{cm})$ & $\rho_{i}/L$ & $\lambda_{\mathrm{mfp}}/L$ \\
 \hline\hline
 solar wind {\small (at $\simc 1~\mathrm{au}$)} & $\simc 10$ & $\simc 0.01$ & $\simc 50$ & $\simc 1$  & $\simc 10^{13}$  & $\simc 10^{-6}$ & $\simc 1$ \\
 \hline
 ICM {\small (at $\simc 100~\mathrm{kpc}$)} & $\simc 10^{-3}$ & $\simc 5$ & $\simc 1$ & $\simc 100$ & $\simc 10^{23}$ & $\simc 10^{-13}$ & $\simc 0.1$ \\
   \hline
 Sgr A* {\small (at $\simc 0.1~\mathrm{pc}$)} & $\simc 100$ & $\simc 2$ & $\simc 10^3$ & $\simc 10$ & $\simc 10^{17}$ & $\simc 10^{-10}$ & $\simc 1$ \\
 \hline
\end{tabular}
\vspace{-1ex}
\caption{Representative order-of-magnitude parameters for the example astrophysical application areas discussed in the text. Here, $n$ is the number density, $T$ is the temperature, $B$ is the magnetic-field strength, $L$ is a macroscopic scale, $\rho_{i}$ is the ion gyroradius (the largest plasma micro-scale at high $\beta$), and $\lambda_{\mathrm{mfp}}$ is the Coulomb mean-free path. Note the enormous disparity between $\rho_i$ and $\lambda_{\mathrm{mfp}} \sim L$.}
\vspace{-2ex}
\label{tab:systems}
\end{table}

\vspace{-2ex}
\paragraph{Black-hole accretion flows.} The theory of black-hole accretion is central to many areas of theoretical, computational, and observational astronomy. Not only does accretion power some of the phenomenologically richest electromagnetic sources in the Universe\cite{Balbus1998}, the flows themselves are excellent laboratories for the study of plasma dynamics and general relativity (GR). Recently, the influence of strong-field GR on black-hole accretion has seen increased attention, with computational efforts to connect the properties of simulated accretion flows in curved spacetime with mm/sub-mm emission observed by the Event Horizon Telescope. These calculations rely upon {\em ad hoc} assumptions about the nature of the accreting plasma, opening an opportunity for those with a knowledge of plasma physics to elucidate the complex interplay between micro-scale plasma processes, meso-scale dynamics (e.g., magnetorotational  turbulence), and macro-scale evolution (e.g., mass accretion, dynamo, particle acceleration).\cite{Kunz2016} Indeed, a key discriminating factor amongst theories of low-luminosity accretion onto the ${\sim}10^6~{\rm M}_\odot$ black hole at our Galactic center, Sgr A$^\ast$, is the ratio of ion-to-electron heating.\cite{Quataert2003} This is an enduring unanswered question in basic plasma physics, complicated by the ${\gtrsim}10^6$ separation between the Schwarzschild radius and the ion gyroscale.

\vspace{-2ex}
\paragraph{The solar wind.} The solar wind is a nearly collisionless plasma of fundamental importance to the Sun--Earth connection. With $\beta\sim 1$ on average, it is not as firmly in the high-$\beta$ regime as the aforementioned astrophysical plasmas.\cite{Bruno2013}
Nonetheless, the solar wind plasma is inhomogeneous, with localized regions at $\beta\gtrsim 1$ where the plasma-kinetic effects discussed above are thought to be fundamental to large-scale plasma dynamics. Indeed, a wide range of dynamically important pressure anisotropies are present in the solar wind, and their extent has been measured to be well constrained by kinetic microinstability thresholds.\cite{Kasper2002,Bale2009} Thus, as well as being influenced by weakly collisional, high-$\beta$ processes, the solar wind is a useful laboratory for studying plasma physics through detailed {\em in situ} spacecraft measurements. 
Such studies can both act as a stringent test of proposed theories and facilitate the discovery of previously unknown plasma processes.

\vspace{-2ex}
\subsection*{Progress and priorities}

In this section, we outline four key areas where progress can be made in understanding the dynamics of weakly collisional, high-$\beta$ plasmas, concluding with suggestions for fruitful paths forward. 

\vspace{-2ex}
\paragraph{Theory.} 
The dynamics of weakly collisional, high-$\beta$ plasmas are governed by  complex, multi-scale interactions between kinetic physics -- gyroscale instabilities driven by significant deviations from local thermodynamic equilibrium -- and large-scale bulk plasma motions. Understanding this interaction on both phenomenological and quantitative levels presents
a fascinating challenge for plasma theory. Such ``basic'' knowledge is also necessary to create a framework for more practical computational and experimental studies (see below). Key areas for study include: (i) further elucidating the most important linear kinetic instabilities (particularly in the presence of strong heat fluxes and kinetic electrons)\cite{Klein2015}; 
(ii) understanding how kinetic instabilities interact with other plasma processes, such as reconnection, heat fluxes, and particle acceleration\cite{Komarov2016,Alt2019,Sironi2015}; and (iii)
understanding the interplay between widely separated fluid macro-scales and kinetic micro-scales.\cite{Squire2019} 

\vspace{-2ex}
\paragraph{Computation.}  

\begin{wrapfigure}[17]{r}[0pt]{0.38\textwidth} 
\vspace{-3.5ex}
\centering
\captionsetup{format=plain}
    \includegraphics[width=0.38\textwidth]{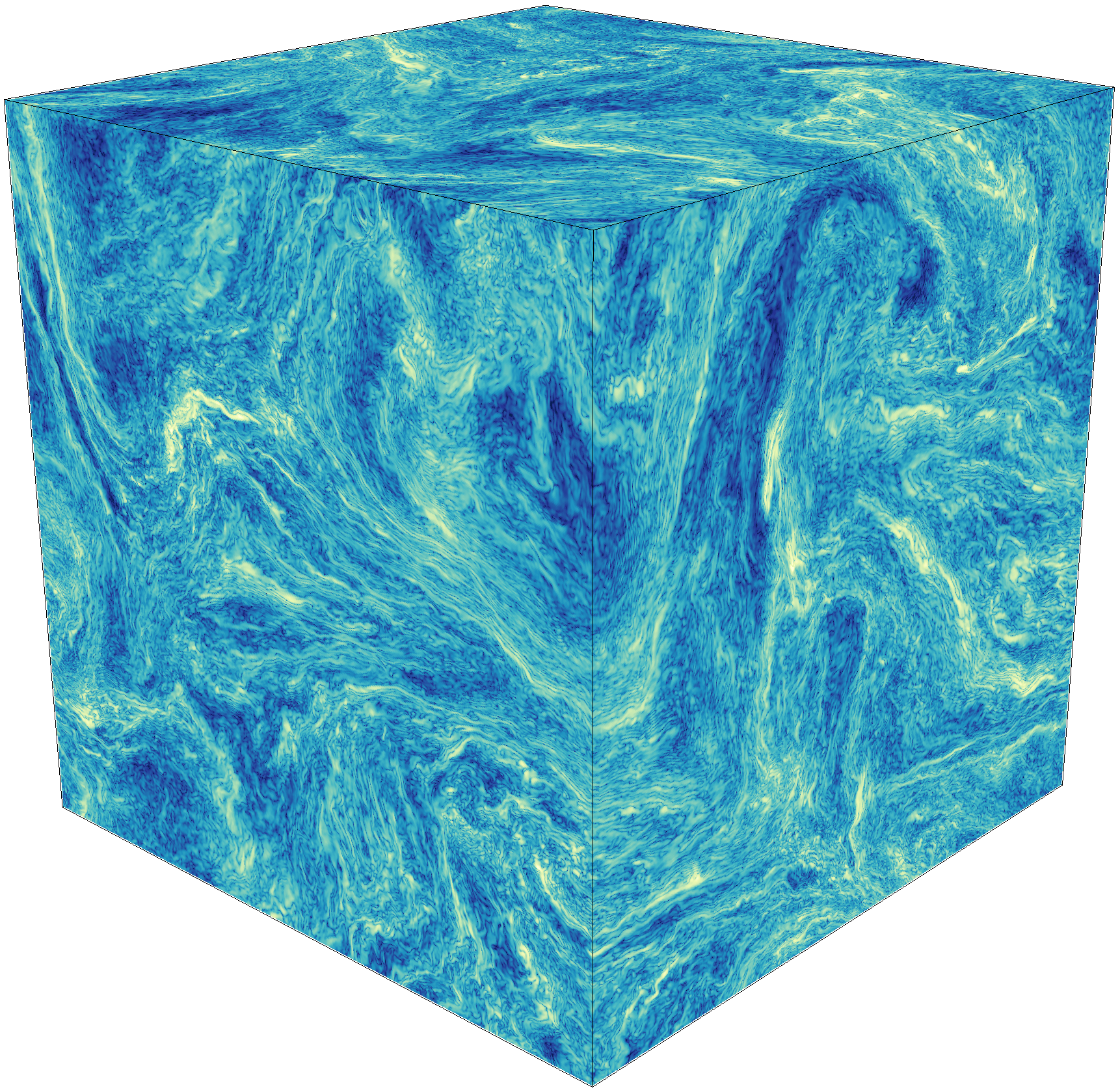}
    \vspace{-4.ex}
    \caption{Plasma dynamo involves complex interactions between turbulent flow and gyroscale instabilities.\cite{St-Onge2018} The magnetic-field strength (shown) and direction affect the flow structure even when $\beta\ggg{1}$.}
    \label{fig:dynamo}
\end{wrapfigure}

Studies of weakly collisional, high-$\beta$ plasmas are computationally challenging, requiring massively parallel computing architectures and highly efficient algorithms. Recent computational advances have enabled 6D, kinetic studies of fluid-scale ($L\gg\rho_{i}$) collisionless systems 
(see Fig.~\ref{fig:dynamo} for an example). These {\em ab initio} approaches, which focus on capturing as much physics as possible, complement the basic theory  
approaches discussed above: theory suggests methods and techniques for simulation analysis; detailed simulations reveal areas where theory may be lacking. Both hybrid-kinetic approaches (with fluid electrons) and fully kinetic approaches have seen fruitful application, as
have 6D continuum and particle-in-cell (PIC) approaches. 

\vspace{-2ex}
\paragraph{In-situ spacecraft and observations.} Our solar system provides an unparalleled laboratory for studying fundamental plasma physics. 
Access to the high-$\beta$ regime is afforded by considering specific spatial regions (e.g., the magnetosheath) or by selecting high-$\beta$ time intervals within larger data sets. New spacecraft (in particular, MMS) are improving kinetic measurements, 
while the accumulation of statistics from long-running older missions is also highly beneficial. Studies of how microinstabilities regulate non-thermal features of the distribution function will inform theory and its application in other astrophysical contexts. Indeed, the analysis of astronomical observations is becoming increasingly dependent on detailed plasma physics: next-generation X-ray telescopes will resolve sub-mean-free-path scales in the ICM, while the Event Horizon Telescope is probing the innermost regions of galactic black-hole accretion flows.

\vspace{-2ex}
\paragraph{Experiments.}
There is no substitute for being able to create a plasma, subject it to stirring, differential rotation, or magnetic fields, and measure its response. Unfortunately, the high-$\beta$ regime is difficult to obtain in the laboratory. Magnetically confined plasmas have $\beta\lesssim{1}$ almost by definition, although the high-$\beta$ regime may be accessible using complex field geometries\cite{bigredball}, helicon sources\cite{HELIX2015}, or mirror configurations. The difficulty of maintaining ion magnetization when $\beta\gtrsim{1}$ suggests that electron-scale instabilities may be more easily studied in the near future.\cite{Cui1992} By contrast, inertially confined plasmas can easily obtain $\beta\gg{1}$ but are necessarily transient. Challenges there include producing plasma hot enough to be weakly collisional and diagnosing such short-lived plasmas.\cite{Huntington2015}

\vspace{-2ex}
\paragraph{\em{What is to be done?}} In light of these four areas of progress, we recommend the following. First, an urgent goal for theory is to formulate and refine fluid models that are simple enough to solve in complex, astrophysically relevant geometries, and detailed  
enough to capture the multi-scale interplay between kinetic gyroscale instabilities, bulk fluid motions, magnetic reconnection, cosmic-ray diffusion, magnetic-field amplification and self-organization, etc. Second, there remain several outstanding algorithmic issues in computation, including a need for stable and accurate methods to reduce particle noise and/or computational expense at high $\beta$, and for methods that resolve important electron kinetics without being limited by the speed of light (e.g., implicit PIC). Notably absent in currently available resources are mid-range (${\sim}10$--$50$M CPU-hour) supercomputing allocations, which are necessary for routine computational studies involving kinetic methods. We thus call for increased availability of mid-range supercomputing options. Finally, there remain many interesting possibilities for future experimental studies of high-$\beta$ plasmas. A program for funding both new university-level plasma experiments and further proposals on existing devices should be prioritized, with accompanying university training of plasma students borne in mind.

\begingroup
\renewcommand{\section}[2]{}%
\bibliographystyle{mnras}

\endgroup

\end{document}